
\def\J{$J/\psi$}
\def\j{J/\psi}
\def\P{$\psi'$}
\def\p{\psi'}

\def\c{c{\bar c}}

\def\t{\tau}

\def\q{q{\bar q}}

\def\lsim{\raise0.3ex\hbox{$<$\kern-0.75em\raise-1.1ex\hbox{$\sim$}}}
\def\gsim{\raise0.3ex\hbox{$>$\kern-0.75em\raise-1.1ex\hbox{$\sim$}}}

\newcount\REFERENCENUMBER\REFERENCENUMBER=0
\def\REF#1{\expandafter\ifx\csname RF#1\endcsname\relax
               \global\advance\REFERENCENUMBER by 1
               \expandafter\xdef\csname RF#1\endcsname
                   {\the\REFERENCENUMBER}\fi}
\def\reftag#1{\expandafter\ifx\csname RF#1\endcsname\relax
               \global\advance\REFERENCENUMBER by 1
               \expandafter\xdef\csname RF#1\endcsname
                      {\the\REFERENCENUMBER}\fi
             \csname RF#1\endcsname\relax}
\def\ref#1{\expandafter\ifx\csname RF#1\endcsname\relax
               \global\advance\REFERENCENUMBER by 1
               \expandafter\xdef\csname RF#1\endcsname
                      {\the\REFERENCENUMBER}\fi
             [\csname RF#1\endcsname]\relax}
\def\refto#1#2{\expandafter\ifx\csname RF#1\endcsname\relax
               \global\advance\REFERENCENUMBER by 1
               \expandafter\xdef\csname RF#1\endcsname
                      {\the\REFERENCENUMBER}\fi
           \expandafter\ifx\csname RF#2\endcsname\relax
               \global\advance\REFERENCENUMBER by 1
               \expandafter\xdef\csname RF#2\endcsname
                      {\the\REFERENCENUMBER}\fi
             [\csname RF#1\endcsname--\csname RF#2\endcsname]\relax}
\def\refs#1#2{\expandafter\ifx\csname RF#1\endcsname\relax
               \global\advance\REFERENCENUMBER by 1
               \expandafter\xdef\csname RF#1\endcsname
                      {\the\REFERENCENUMBER}\fi
           \expandafter\ifx\csname RF#2\endcsname\relax
               \global\advance\REFERENCENUMBER by 1
               \expandafter\xdef\csname RF#2\endcsname
                      {\the\REFERENCENUMBER}\fi
            [\csname RF#1\endcsname,\csname RF#2\endcsname]\relax}
\def\refss#1#2#3{\expandafter\ifx\csname RF#1\endcsname\relax
               \global\advance\REFERENCENUMBER by 1
               \expandafter\xdef\csname RF#1\endcsname
                      {\the\REFERENCENUMBER}\fi
           \expandafter\ifx\csname RF#2\endcsname\relax
               \global\advance\REFERENCENUMBER by 1
               \expandafter\xdef\csname RF#2\endcsname
                      {\the\REFERENCENUMBER}\fi
           \expandafter\ifx\csname RF#3\endcsname\relax
               \global\advance\REFERENCENUMBER by 1
               \expandafter\xdef\csname RF#3\endcsname
                      {\the\REFERENCENUMBER}\fi
[\csname RF#1\endcsname,\csname RF#2\endcsname,\csname
RF#3\endcsname]\relax}
\def\refand#1#2{\expandafter\ifx\csname RF#1\endcsname\relax
               \global\advance\REFERENCENUMBER by 1
               \expandafter\xdef\csname RF#1\endcsname
                      {\the\REFERENCENUMBER}\fi
           \expandafter\ifx\csname RF#2\endcsname\relax
               \global\advance\REFERENCENUMBER by 1
               \expandafter\xdef\csname RF#2\endcsname
                      {\the\REFERENCENUMBER}\fi
            [\csname RF#1\endcsname,\csname RF#2\endcsname]\relax}
\def\Ref#1{\expandafter\ifx\csname RF#1\endcsname\relax
               \global\advance\REFERENCENUMBER by 1
               \expandafter\xdef\csname RF#1\endcsname
                      {\the\REFERENCENUMBER}\fi
             [\csname RF#1\endcsname]\relax}
\def\Refto#1#2{\expandafter\ifx\csname RF#1\endcsname\relax
               \global\advance\REFERENCENUMBER by 1
               \expandafter\xdef\csname RF#1\endcsname
                      {\the\REFERENCENUMBER}\fi
           \expandafter\ifx\csname RF#2\endcsname\relax
               \global\advance\REFERENCENUMBER by 1
               \expandafter\xdef\csname RF#2\endcsname
                      {\the\REFERENCENUMBER}\fi
            [\csname RF#1\endcsname--\csname RF#2\endcsname]\relax}
\def\Refand#1#2{\expandafter\ifx\csname RF#1\endcsname\relax
               \global\advance\REFERENCENUMBER by 1
               \expandafter\xdef\csname RF#1\endcsname
                      {\the\REFERENCENUMBER}\fi
           \expandafter\ifx\csname RF#2\endcsname\relax
               \global\advance\REFERENCENUMBER by 1
               \expandafter\xdef\csname RF#2\endcsname
                      {\the\REFERENCENUMBER}\fi
        [\csname RF#1\endcsname,\csname RF#2\endcsname]\relax}
\def\refadd#1{\expandafter\ifx\csname RF#1\endcsname\relax
               \global\advance\REFERENCENUMBER by 1
               \expandafter\xdef\csname RF#1\endcsname
                      {\the\REFERENCENUMBER}\fi \relax}

%

\def\NP{{ Nucl.\ Phys.\ }}
\def\PL{{ Phys.\ Lett.\ }}

\def\PRL{{ Phys.\ Rev.\ Lett.\ }}

\def\ZP{{ Z.\ Phys.\ }}

\def\etal{{\sl et al.}}

\magnification=1200
\hsize=16.0truecm
\vsize=23.0truecm
\baselineskip=13pt
\pageno=0
\def\la{\Lambda_{\rm QCD}}

\def\q{\q{\bar q}}

\hfill CERN-TH/95-73
\par\hfill BI-TP 95/16
\vskip 2 truecm
\centerline{\bf CHARMONIUM INTERACTION IN NUCLEAR MATTER}
\vskip 1.5 truecm
\centerline{\bf D. Kharzeev and H. Satz}
\bigskip
\centerline{Theory Division, CERN}
\centerline{CH-1211 Geneva, Switzerland}
\medskip
\centerline{and}
\medskip
\centerline{Fakult\"at f\"ur Physik, Universit\"at Bielefeld}
\centerline{D-33501 Bielefeld, Germany}
\vskip 2 truecm
\centerline{\bf Abstract:}
\medskip
We analyse the kinematic regimes attainable for \J~and \P~production
in $p-A$ \break  collisions. With this information, we specify the
requirements for an experiment to study the interaction of physical
charmonium states in nuclear matter, making use of a heavy ion
beam incident on a hydrogen or deuterium target.
\par \vfill
\noindent CERN-TH/95-73
\par\noindent
BI-TP 95/16
\par\noindent
April 1995\par
\eject
{}~~~\par\vfill
\noindent {\bf 1. Introduction}
\medskip
The production of charmonium states in hadron-nucleus and
nucleus-nucleus collisions provides an excellent tool to study
different aspects of strongly interacting matter. At high energy,
charmonium production proceeds through gluon fusion to a colour octet
$\c$ state (Fig.\ 1a), which subsequently
neutralises its colour and obtains the right quantum numbers by
interacting with the colour field of the collision \ref{quarko}.
At lower energies, quark-antiquark annihilation adds significant
contributions (Fig.\ 1b), which for large $|x_F|$ become dominant. In
both parton processes, the first production stage is a colour octet $\c$
state which needs a certain time before it becomes colour neutral.
Consider $p-A$ collisions leading to the production of charmonia fast in
the rest frame of the target nucleus. The nuclear medium then sees
only the passage of a fast coloured $\c$ pair, with which it interacts
as with any colour charge. There is nothing charmonium-specific about
this interaction, and in particular, the nucleus does not know what
hadronic state this pair will turn into later on. To study the effect of
the medium on the different physical charmonium states, these have to be
sufficiently slow to be fully formed inside the medium. In the first
part of this note, we will identify the kinematic regions
corresponding to these different situations. As we shall
show, all hadron-nucleus collisions studied so far provide
information only about the passage of the coloured $\c$ pair through
nuclear matter; nothing is known experimentally about the interaction of
fully formed charmonia with the nuclear medium.
\par
This lack of information is particularly serious if one wants to use
charmonium production as a probe for deconfinement in nucleus-nucleus
collisions. It was predicted \ref{Matsui} that quark-gluon plasma
formation leads to \J~suppression in nuclear collisions, and such a
suppression was subsequently in fact observed \ref{Baglin}. This
observation in turn triggered a ``conventional" explanation
\ref{absorb}, in which the suppression was attributed to absorption of
the \J~in a dense hadronic (and hence confined) medium. More recently it
was shown \ref{KS3} that the theory of heavy quarkonium interactions
with light hadrons leads to a break-up cross section which rules out
such an absorption. It was also noted, however, that a direct
experimental test of the crucial cross section behaviour is still
lacking. The theory on which it is based becomes exact in the heavy
quark limit, and although some related checks were favorable, it is
certainly necessary to verify experimentally that the charm quark mass
is already sufficiently heavy. In the second part of this note, we
compare the predictions of heavy quark QCD with previous geometric
estimates, in which the QCD threshold behaviour had not been taken into
account. From this comparison it becomes clear that an experimental
verification of the predicted \J-nucleon cross section behaviour is
possible. After a discussion of possible initial state modifications
(EMC effect) we then specify the kinematic range and further aspects of
an experiment specifically devoted to this question.
\par\eject
\noindent
{\bf 2. Production Kinematics}
\medskip
The time needed by the colour octet $\c$ state to become colour neutral
can be estimated in terms of the energy of a further gluon
``evaporated" to produce a colour singlet state \ref{KS1}. In the
rest frame of the $\c$, this time $\t_0$ becomes
$$
\t_0 = {1\over \sqrt{2m_c \omega}}; \eqno (1)
$$
$m_c$ is the charm quark mass and $\omega$ the gluon energy. In confined
matter, $\omega~\gsim~\la$, making the colour neutralisation time in
the rest frame of the $\c$ pair $\t_0 \simeq 0.25$ fm. In the rest
frame of the nucleus, the $\c$ travels in the time $\t_0$ the distance
$$
d_0 = \left({P_A\over M}\right) \t_0, \eqno(2)
$$
with $P_A$ denoting the momentum and $M$ the mass of the $\c$ state.
{}From Eq.\ (2) it is thus clear that sufficiently fast $\c$ are still
coloured when they leave the nuclear environment.
\par
The Feynman variable $x_F$ of the charmonium state and hence of the
corresponding $\c$ is
$$
x_F \equiv \left( {P\over P_{max}} \right), \eqno(3)
$$
where
$$
P = \gamma P_A - \gamma \beta \sqrt{P_A^2+M^2}  \eqno(4)
$$
is the center of mass (cms) momentum of the $\c$; $\gamma$ and $\beta$
specify the transformation from the nuclear rest frame to the cms.
The maximum cms momentum the $\c$ can have is
determined by the relation
$$
s = (\sqrt{P_{max}^2+M^2} + \sqrt{P_{max}^2 + 4m^2})^2, \eqno(5)
$$
with $s$ for the squared incident cms energy and $m$ for
the nucleon mass; Eq.\ (5) corresponds to the configuration of both
nucleons going forward and the $\c$ backward.
\par
We shall now consider the specific case of a 160 GeV/c proton beam
incident on a nuclear target, corresponding to $\sqrt{ s} = 17.4$
GeV. At this energy, it is possible to study $p-A$ collisions at the
CERN-SPS with a nuclear beam incident on a hydrogen target as well as
the conventional inverse.
In Table 1, we list the resulting windows for the production of \J,
$\chi_c$ and \P. We thus obtain 8.3 GeV as the maximum momentum
which a \J~can have in the cms; this means that in the nuclear rest
frame, it has a minimum momentum of 4.7 GeV. In terms of rapidity, the
\J~production range at this energy is $|y_{\rm cms}| \leq 1.7$; in the
region between $|y_{\rm cms}|=1.7$ and the maximum rapidity
$|y_{\rm cms}|=2.9$, \J~production is kinematically not possible.
\par
Since we want to study the interaction of physical charmonium states
with the nuclear medium, we want to avoid the kinematic regime in which
the $\c$ can interact with the nucleus already in its colour octet
phase. Hence we require $d_0 \leq$ 1.5 fm, so that the $\c$
has become colourless before it leaves the range of the nucleon on which
it was produced. Using the above relations, this implies at
the noted collision energy $x_F~\lsim  -0.2$; for all larger $x_F$, at
least a partial colour interaction with the nuclear medium remains. In
the region $-0.2 \leq x_F \leq 0$, this is effectively negligible, since
the distance the $\c$ travels as coloured state is only about 2 fm at
$x_F=0$. The colour phase increases rapidly with increasing $x_F$, and
for $x_F \gsim ~0.2$, the $\c$ is coloured on its entire path through
the nucleus \ref{KS1}. Hence the nuclear effects on \J~and
\P~observed in this region must be identical. As already mentioned, all
presently available data \refs{Badier}{E772} on charmonium production in
hadron-nucleus collisons fall into this region, and they in fact observe
the same nuclear suppression of \J~and \P~production.
\par
If we wish to study the interaction of fully formed physical resonances
with a nuclear medium, it is clear then that we must require $x_F\leq
-0.2$; this is a necessary, not a sufficient condition. For the latter,
we have to assure that the $\c$ has not only lost its colour, but that
it has also had sufficient time to attain its full physical size.
An estimate of the resonance formation time $\t_r$ for the \J~is
obtained from potential theory \ref{Karsch}, with $\t_r(\j)\simeq
0.35$ fm; this agrees with the rather model-independent upper bound
based on the uncertainty relation and the separation between ground
state and first excited state. The $\chi_c$ and \P~values are larger;
the noted potential theory studies are in accord with about 1 fm
for both. With these values and the relation
$$
d_r = \left({P_A \over M}\right) \t_r    , \eqno(6)
$$
we find that the \J~passes through the entire nuclear medium as a fully
formed resonance for $x_F\lsim -0.45$, but that at the energy under
consideration the $\chi_c$ and \P~never reach that stage. Although
they become colourless in the medium when $x_F\leq -0.2$, they have not
quite reached full physical size when leaving the medium even for
$x_f\simeq -1$.
\bigskip
\noindent
{\bf 3. Charmonium Absorption}
\medskip
The effect of the nuclear medium on the passing coloured $\c$ state has
been measured at two different energies \refs{Badier}{E772}; it can be
understood in terms of the energy loss of the colour octet in a confined
medium \ref{KS1} together with quantum-mechanical coherence effects
(nuclear shadowing) \ref{KS2}. The results are schematically
illustrated in Fig.\ 2 in the range $x_F\geq 0.1$; as noted, the
effects on \J~, $\chi_c$
and \P~are identical. Below $x_F=0$, the $\c$ passes the nucleus in
part as a colourless state, and below $x_F\simeq -0.2$, it is
colourless
on its entire path through the nucleus. We can parametrise the survival
probability for state $i$ ($i=\j,~\chi_c,~\p$) as
$$
S_i = exp\{-n_0 \sigma_i L\}, \eqno(7)
$$
where $n_0=0.17$ fm$^{-3}$ is the normal nuclear density,
$L=(3/4)R_A = (3/4)1.15~A^{1/3}$ the average path length in the nucleus
and $\sigma_i$ the absorption cross section of the state $i$ in nuclear
matter.\footnote{*}{We note here that if the form (7) is used to
parametrise the nuclear suppression of charmonium production at
positive $x_F$ \ref{Gerschel}, then $\sigma_i$ becomes the
$i$-independent cross section
for the interaction of a colour octet $\c$ with a nucleon.
This can be quite large, but it is not related to that for
the interaction of physical quarkonium states with nucleons.}
If the charmonium state is fully formed before it leaves the
range of the nucleon at which it was produced, $\sigma_i$ is simply
the charmonium-nucleon cross section. If it is not yet fully formed, the
effective cross section will be smaller, vanishing in the colour
transparency limit of a pointlike colour singlet $\c$ state. In
principle, the interaction of this evolving resonance should be studied
quantum-mechanically. Here we shall for simplicity parametrise the
cross section as function of the distance $d$ which the state has
travelled \refs{Farrar}{Blaizot}, so that
$$
\sigma_i(d) = \sigma_i \left( {d\over {\bar L_i}}\right)^2, \eqno(8)
$$
where $\sigma_i$ is the fully developed cross section and ${\bar
L}=[(P_A/M_i)\t_r^{(i)} - 1]$ the effective distance travelled until
full resonance
formation. Eq.\ (8) holds for $d\leq {\bar L}$; for $d\geq {\bar L}$,
$\sigma_i(d)=\sigma_i$. Using this parametrisation, Eq.\ (7) is
replaced by
$$
S_i = exp\{-n_0\sigma_i[L - {2\over 3} {\bar L_i}]\} \eqno(9)
$$
whenever  $d \leq {\bar L}$. To determine the actual survival
probabilities, we now still need the cross sections $\sigma_i$ for
the fully formed resonances colliding with nucleons.
\par
\refadd{Bhanot}
\refadd{SVZ}
\refadd{Kaidalov}
In the geometric approach to charmonium absorption, $\sigma_i$ is
assumed to be the total {\sl high energy} collision cross section. This
can be estimated by geometric arguments \ref{Povh}, giving
$\sigma_{\j}\simeq 2.5$ mb, $\sigma_{\chi_c}\simeq 6.0$ mb and
$\sigma_{\p}\simeq 9.2$ mb. With this, the picture
is complete: for each value of $x_F$, we have the corresponding
momentum $P_A$ in the nuclear rest frame; from that we get in turn the
distance travelled in the medium as nascent resonance and the associated
cross section. The difference in asymptotic cross sections and the
different formation times then lead to the absorption patterns shown in
Fig.\ 2 for negative $x_F$. The region around $x_F=0$ is drawn as
continous interpolation of the forms at positive and negative $x_F$.
\par
The crucial assumption leading to the behaviour just discussed is that
the cross sections for the interaction of charmonium states with
nucleons attain their full asymptotic values at threshold. The
invariant energy $\sqrt {\bar s}$ for the quarkonium-nucleon
interaction is below 8 - 9 GeV for the region of $x_F \leq 0$; for
$x_F \leq -0.5$, $\sqrt {\bar s} \leq $ 6 GeV. The threshold for
$D{\bar D}$ production in charmonium-nucleon interactions is 4.7 GeV.
Hence it is the inelastic cross section {\sl near threshold} that
matters for the survival probability at negative $x_F$. For the \J,
and perhaps also for the $\chi_c$, this cross section is
calculable by short distance QCD \refto{Bhanot}{Kaidalov}\ref{KS3}. A
break-up requires the interaction with hard gluons, but gluons
confined to slow hadrons in the charmonium rest frame are generally
very soft. As a consequence, the break-up cross
sections near threshold are expected to be much smaller than their
asymptotic values. For the \J, the heavy quark theory predicts
\ref{KS3}
$$
\sigma_{\j N}({\bar s}) \simeq 2.5~{\rm mb} \times \left( 1 - \left[{
2M_{\j}(m+\epsilon_{\j}) \over ({\bar s}- M_{\j}^2)}\right]
\right)^{6.5},
\eqno (10)
$$
where $\epsilon_{\j}=2M_D-M_{\j}\simeq 0.64$ GeV is the \J~binding
energy and $m$ the nucleon mass. The behaviour of this cross section is
shown in Fig.\ 3 as function of the momentum $P_N$ of a nucleon incident
on a \J~at rest. For $P_N \simeq 4$ GeV/c, corresponding to
$\sqrt{\bar s} = 6$ GeV, it is almost two orders of
magnitude below its asymptotic value. For a cross section of
this size, the survival probability is in good approximation unity, and
that is the basis of the prediction that confined matter is transparent
for \J's.
\par
The theoretical basis for the use of short distance QCD is much
less reliable for the $\chi_c$, since here the binding energy is
just around $\la$. Keeping this in mind, we shall nevertheless see what
a corresponding analysis leads to. Instead of Eq.\ (10) we now obtain
$$
\sigma_{\chi N}({\bar s}) \simeq 11.3~{\rm mb} \times \left( 1 -
\left[{
2M_{\chi}(m+\epsilon_{\chi}) \over ({\bar s}- M_{\chi}^2)}\right]
\right)^{6.5},
\eqno (11)
$$
where $\epsilon_{\chi}=2M_D-M_{\chi}\simeq 0.24$ GeV now denotes the
binding energy of the $\chi_c$. The asymptotic value is a factor two
larger than the geometric estimate; this is a consequence of the fact
that short distance QCD \refs{Bhanot}{SVZ} leads to higher
powers in the bound state radii than just $r^2$. The behaviour of the
$\chi_c  N$ cross section (11) is also shown in Fig.\ 3.
\par
The \P~state lies essentially at the open charm threshold and hence
has a binding energy much less than $\la$; it is therefore
definitely not calculable in short distance QCD. The asymptotic
form \P$_{\infty}$
shown in Fig.\ 2 may thus be a reasonable estimate here.
\par
The resulting survival probabilies $S_{\j}(x_F)$ and $S_{\chi}(x_F)$
are compared in Fig.\ 2 to those obtained above in the geometric
approach. In the region around $x_F=0$ we again make an
estimate taking into account both the vanishing of
the colour interaction with decreasing $x_F$ and the growing
charmonium-nucleon cross section with increasing $x_F$.
\bigskip
\noindent
{\bf 4. EMC Effect}
\medskip
The survival probabilities shown in  Fig.\ 2 are related to the measured
$p-A$ and $p-p$ production cross sections through
$$
{(d\sigma_i^{pA}/dx_F)\over A(d\sigma_i^{pp}/dx_F)} = {g_A(x_2)\over
g_p(x_2)} S_i(x_F) \equiv R_{A/p}(x_2) S_i(x_F), \eqno(12)
$$
where $g_A(x_2)$ and $g_p(x_2)$ are the parton distribution functions in
nuclear and proton target, respectively. The fractional parton momentum
$x_2$ is given by
$$
x_2 = {1\over 2}(\sqrt{(x_F^2 + (4M_i^2/s))} \pm x_F) \eqno(13)
$$
in terms of the variables $x_F$ and $s$; the plus sign holds for
positive, the minus sign for negative $x_F$. At the energy under
discussion above (${\sqrt s}=17.4$ GeV) and for $x_F\leq 0$, we have
$x_2\geq 0.15$; we are thus above the region in which quantum-mechanical
coherence effects (nuclear shadowing or antishadowing) play a role
\ref{KS2}.
However, we know from deep inelastic scattering on nuclear targets that
for $x_2\geq 0.15$ the quark parton distributions in nuclei are
modified in comparison to those in nucleons. This modification,
generally denoted as EMC effect \ref{EMC}, is shown in Fig.\ 4.
\par
Although the EMC effect has so far been observed only for quarks, it is
to be expected that gluons will exhibit a similar behaviour, so that
the initial state factor $R_{A/p}(x_2)$ will introduce a
further $x_F$ variation in addition to that coming from $S_i(x_F)$.
With increasing $|x_F|$, charmonium production is more and more due to
quark-antiquark annihilation rather than to gluon fusion;
the two contributions become approximately equal around $|x_F|=0.5$,
and as $|x_F| \to 1$, the $q{\bar q}$ contribution is dominant
\refs{quarko}{Ramona}. We shall here assume that quark and gluon
distributions behave in the same way and use the quark form of
$R_{A/p}$ in the whole region $-1.0 \leq x_F\leq 0$; this gives us a
prediction for the behaviour of the cross section ratio in that region.
It is illustrated in Fig.\ 5 for the directly produced \J~state in
the two cases considered.
\par
As a cross check of the form of $R_{A/p}(x_2)$, the EMC effect in this
region can also be studied independently. Measuring Drell-Yan
dilepton production there provides $R_{A/p}(x_2)$ for quarks
directly, without any final state modification. A measurement of open
charm production leads to $R_{A/p}(x_2)$ in just the same
superposition of quark-antiquark annihilation and gluon fusion as in
charmonium production, but again without any final state effect.
Separate measurements of Drell-Yan and/or open charm production would
thus determine the EMC modification without addititional final state
effects. Such measurements can therefore be used to remove the EMC
modification of \J~and \P~production data, which can then be compared
directly to the predictions shown in Fig.\ 2.
\par
Such measurements would moreover be of considerable interest in
themselves. At positive $x_F$, little or no nuclear modifications are
seen in Drell-Yan production \ref{E772-DY}. The same experiment also
showed no nuclear effects on open charm production. These two facts are
part of the empirical basis for the claim \ref{KS2} that the
considerable suppression of charmonium production in $p-A$ collisions
at positive $x_F$ is due to quantum-mechanical coherence effects
and energy loss suffered by the passing virtual colour charge, and not
to factorisable parton distribution function changes. The measurement
of factorisable modifications for negative $x_F$ would provide further
support to this interpretation.
\bigskip
\noindent
{\bf 5. Experimental Aspects}
\medskip
The study of charmonia or Drell-Yan dileptons slow in the nuclear rest
frame has up to now been essentially impossible. Both require the
detection of slow dileptons, and for this the abundance of slow hadrons
constitutes an overwhelming background. In the case of
fast dileptons, a hadron absorber can eliminate these, and hence all
$p-A$ studies were restricted to dilepton pairs of more than 20 GeV in
the rest frame of the nuclear target. This in turn restricts the
available data to $x_F \gsim~0$.
\par
The advent of the $Pb$-beam at the CERN-SPS has removed this constraint.
With the $Pb$-beam incident on a hydrogen (or deuterium) target, the
nuclear rest frame moves with a lab momentum of 160 GeV.
Hence now charmonia and their decay dileptons are very fast in the lab
system and will thus pass the hadron absorber. The window for such
measurements is evident from Table 1, and the expected behaviour for
positive $x_F$ is that shown in Figs.\  2 for negative $x_F$. For
clarity, we denote the variables in the case of a $A$-beam incident on a
hydrogen target by a superscript $A$. Thus the region of greatest
interest is $0.4 \geq x_F^A \geq 1$; in terms of rapidity, this
corresponds to the range $3.6 \geq y_{\rm lab}^A \geq 4.5$.
\par
The above predictions for \J~production correspond to directly
produced 1S $\c$ resonances. The \J~peak observed in the measured
dilepton spectrum is about 60\% due to this origin and about 40\%
due to direct $\chi_c$ production with subsequent decay $\chi_c \to
\j + \gamma$ (see \ref{quarko}). Unless it is possible to measure
$\chi_c$ production
independently \ref{Anton}, the data will contain a superposition of
direct $1S$ \J~production and \J~'s from $\chi_c$
decay.\footnote{*}{In addition, a few percent ($\sim$ 5 \%) will
come from
\P~decays. We neglect this in both scenarios.} To estimate the
effect of this in the two scenarios considered here, we simply add the
corresponding predictions with the noted 60/40 weights; the result is
shown in Fig.\ 6.
As seen, for $x_F\leq -0.4$, the two approaches differ qualitatively
in their functional form and quantitatively by more than 20\%. An
experimental test should therefore be possible.
\par
Finally we note that the analysis proposed here is a comparison of
production data from a heavy nuclear beam on a hydrogen or deuterium
target with that from a proton beam on the same target. Although there
are measurements for the latter (see \ref{quarko} for a compilation), it
seems very desirable to obtain both $A$ and $p$ beam data in the same
experiment, in order to avoid acceptance uncertainties.
\bigskip
\noindent
{\bf 6. Conclusions}
\medskip
The study of charmonium production at low momenta in the nuclear rest
frame, together with that of Drell-Yan dileptons and open charm,
opens a completely unexplored region of the behaviour of hard probes in
nuclear matter. Such studies have become experimentaly feasible only
with the advent of the $Pb$-beam at the CERN-SPS, and since they
require very forward measurements, they would not be easy to carry out
in future collider experiments. The results of such a program would have
a decisive impact on at least three different topics:
\item{--}{They would test directly the heavy quark theory prediction
of strong threshold damping for the interaction of \J's with light
hadrons. This is crucial for the use of charmonium production as
deconfinement probe.}
\item{--}{They would allow a study of the EMC effect by $\c$
production and thus provide a complementary tool to deep inelastic
scattering on nuclear targets. This is important for the investigation
of nuclear modifications of parton distributions.}
\item{--}{By comparing nuclear effects on charmonia with those on
Drell-Yan production, the role of modified parton
distributions could be tested directly. Fast $\c$ pairs (in the nuclear
rest frame) show a very different behaviour than fast dileptons,
and this rules out the factorisable modification of
parton distributions there. For slow $\c$ and dileptons, the behaviour
is predicted to be similar and and the modifications factorisable.}
\par\noindent
The experimental investigation of $Pb-p$ collisions could thus do much
to further our understanding of the effect of a confined (nuclear)
medium on the parton structure of hadronic systems.
\bigskip\noindent
{\bf Acknowledgement}
\medskip
The financial support of the German Research Ministry (BMFT) under
Contract 06 BI 721 is gratefully acknowledged.
\vfill\eject
\centerline
{\bf References}
\medskip\parindent=14pt
\item{\reftag{quarko})}{R. V. Gavai et al., ``Quarkonium
Production in Hadronic Collisions", in {\sl
Hard Processes in Hadronic Interactions}, H. Satz and X.-N. Wang
(Eds.); CERN Preprint CERN-TH.7526/94 (December 1994).}
\par
\item{\reftag{Matsui})}{T. Matsui and H. Satz, \PL B 178 (1986)
416.}
\par
\item{\reftag{Baglin})}{C. Baglin et al., \PL B220 (1989) 471; B251
(1990) 465, 472; B225 (1991) 459.}
\par
\item{\reftag{absorb})}{
See e.g., J.-P. Blaizot and J.-Y. Ollitrault,
in {\sl Quark-Gluon Plasma}, R. C. Hwa (Ed.), World Scientific,
Singapore 1990.})

\par
\item{\reftag{KS3})}{D. Kharzeev and H. Satz, \PL B 334 (1994) 155.}
\par
\item{\reftag{KS1})}{D. Kharzeev and H. Satz, \ZP C 60 (1993) 389.}
\par
\item{\reftag{Badier})}{J. Badier et al., \ZP C 20 (1983) 101.)
\par
\item{\reftag{E772})}{D. M. Alde et al., \PRL 66 (1991) 133.}
\par
\item{\reftag{Karsch})}{F. Karsch and H. Satz, \ZP 51 (1991) 209.}
\par
\item{\reftag{KS2})}{D. Kharzeev and H. Satz, \PL B 327 (1994) 361.}
\par
\item{\reftag{Gerschel})}{C. Gerschel and J. H\"ufner, \ZP C 56 (1992)
171.}
\par
\item{\reftag{Farrar})}{G. Farrar et al., \PRL 64 (1990) 2996.}
\par
\item{\reftag{Blaizot})}{J.-P. Blaizot and J.-Y. Ollitrault, \PL 217B
(1989) 386.}
\par
\item{\reftag{Bhanot})}{G. Bhanot and M. E. Peskin, \NP B 156 (1979)
391.}
\par
\item{\reftag{SVZ})}{M. A. Shifman, A. I. Vainshtein and V. I.
Zakharov, \PL 65B (1976) 255.\hfill \break
V. A. Novikov, M. A. Shifman, A. I. Vainshtein and V. I. Zakharov,
\NP B136 (1978) 125.}
\par
\item{\reftag{Kaidalov})}{A. Kaidalov, in {\sl QCD and High Energy
Hadronic Interactions}, J. Tran Thanh Van (Ed.), Ed. Frontieres,
Gif-sur-Yvette 1993.}
\par
\item{\reftag{Povh})}{J. H\"ufner and B. Povh, 58 (1987) 1612.}
\par
\item{\reftag{EMC})}{
J.~J.~Aubert \etal, \PL B123 (1983)275;\hfill\break
A.~C.~Benvenuti \etal, \PL B189 (1987) 483;\hfill\break
P.~Amaudruz \etal, \ZP C51 (1991) 387.}
\par
\item{\reftag{Ramona})}{R. Vogt, private communication.}
\par
\item{\reftag{E772-DY})}{{D. M. Alde et al., \PRL 66 (1991) 2285.}
\par
\item{\reftag{Anton})}{{L. Antoniazzi et al., \PRL 70 (1993) 383.}
\par
\vfill\eject
{}~~~\medskip
\centerline{\bf Table 1:}
\medskip
\centerline{Windows for Charmonium Production in $p-A$ Collisions at
$\sqrt{s}$ = 17.4 GeV}
\medskip
$$
{\offinterlineskip \tabskip=0pt
\vbox{
\halign to 0.9\hsize
{\strut
\vrule width0.8pt\quad#
\tabskip=0pt plus100pt
& # \quad
&\vrule#&
&\quad \hfil # \quad
&\vrule#
&\quad \hfil # \quad
&\vrule#
&\quad \hfil # \quad
\tabskip=0pt
&\vrule width0.8pt#
\cr
\noalign{\hrule}\noalign{\hrule}
&~~~~~~~~~ &&~~~~~~~~~~~~~&&~~~~~~~~~~~~~~ &&~~          ~~~~~~~&\cr
& Variable &&     \J      &&     $\chi_c$  &&       \P          &\cr
&~~~~~~    &&~~~~~~~~~~~~~&&~~~~~~~~~~~~~~ &&              ~~~~~&\cr
\noalign{\hrule}
&~~        &&~~           &&               &&                   &\cr
& $|P^{\rm max}_{\rm CMS}|$~[GeV]
           &&      8.31   &&  8.24         &&  8.19             &\cr
&~~        &&~~           &&~~             &&~                  &\cr
\noalign{\hrule}
&          &&             &&               &&                   &\cr
& $|y^{\rm max}_{\rm CMS}|$
           &&       1.71  &&         1.59  &&  1.54             &\cr
&~~        &&~~           &&~~             &&~                  &\cr
\noalign{\hrule}
&~~        &&~~           &&               &&                   &\cr
& $~P^{\rm min}_A~$~[GeV]
           &&       4.70  &&         6.12  &&  6.89             &\cr
&~~        &&~~           &&~~             &&~                  &\cr
\noalign{\hrule}
&          &&             &&               &&                   &\cr
& $ y^{\rm min}_A $
           &&       1.21  &&         1.33  &&  1.38             &\cr
&~~        &&~~           &&~~             &&~                  &\cr
\noalign{\hrule}
&          &&             &&               &&                   &\cr
& $ y^{\rm max}_{\rm lab} $
           &&       4.63  &&         4.51  &&  4.45             &\cr
&~~        &&~~           &&~~             &&~                  &\cr
\noalign{\hrule}}
}}
$$
\vfill\eject
{}~~~\medskip
\centerline{\bf Figure Captions}
\medskip
\parindent=0pt
Fig.\ 1: \J~production by gluon fusion (a) and by quark-antiquark
annihilation (b).
\medskip
Fig.\ 2: Charmonium production in $p-Pb$ collisions at 160 GeV incident
beam energy.  The suppression for $x_F\geq 0$ is the same for \J, \P~
and $\chi_c$; data are for \J~production at 200 GeV beam energy
\ref{Badier}. The predicted suppression for $x_F \leq 0$ is shown for
asymptotic cross sections (solid lines) and for cross sections from
short distance QCD (dashed lines).
\medskip
Fig.\ 3: Dissociation cross sections for \J-nucleon and $\chi_c$-nucleon
interactions as function of the momentum $P_N$ of a nucleon incident
on a charmonium at rest.
\medskip
Fig.\ 4: EMC effect: ratio $R_{A/p}=g_A(x)/g_p(x)$ of quark
distribution function $g_A(x)$ in heavy nuclei to that in nucleons
$g_p(x)$ \ref{EMC}.
\medskip
Fig.\ 5: Direct \J~production in $p-Pb$ collisions in the geometric
approach and from short distance QCD, with (solid line) and without
(dashed line) EMC effect modification.
\medskip
Fig.\ 6: \J~production in $p-Pb$ collisions in the geometric scenario
and in short distance QCD, with 40\% of the observed \J~coming from
$\chi_c$ decay; without EMC effect.

\vfill\bye